\documentclass[12pt]{iopart}

\usepackage{color}
\usepackage{scalerel}
\usepackage{subeqnarray}
\usepackage{amssymb}
\usepackage{graphicx}
\usepackage{bm}
\usepackage{hyperref}
\usepackage{float}
\usepackage{mathrsfs}

\newcommand\sh{\stretchrel*{$/$}{\textsc{e}}}
\newcommand{\red}{\color{red}}

\def\be{\begin{equation}}
\def\ee{\end{equation}}

\begin{document}

\title[Physics of rowing oars]{Physics of rowing oars}

\author{Romain Labb\'e$^{1,2}$, Jean-Philippe Boucher$^{1,2}$, Christophe Clanet$^1$ and Michael Benzaquen$^1$ \medskip}
\address{$^1$ LadHyX, UMR CNRS 7646, Ecole polytechnique, 91128 Palaiseau Cedex, France}
\address{$^2$ Phyling, 73 rue LŽon Bourgeois, 91120 Palaiseau, France}
\ead{michael.benzaquen@polytechnique.edu}
\begin{abstract}
In each rowing sport, the oars have their very own characteristics most of the time selected through a long time experience. Here we address experimentally and theoretically the problem of rowing efficiency as function of row lengths and blades sizes. In contrast with previous studies which consider imposed kinematics, we set an imposed force framework which is closer to human constraints. 
We find that optimal row lengths and blades sizes depend on sports and athletes strength, and we provide an optimisation scheme. 
\end{abstract}

\vspace{2pc}

\section*{Introduction}

\begin{figure}[t]
\centerline{\includegraphics[width=0.6\columnwidth]{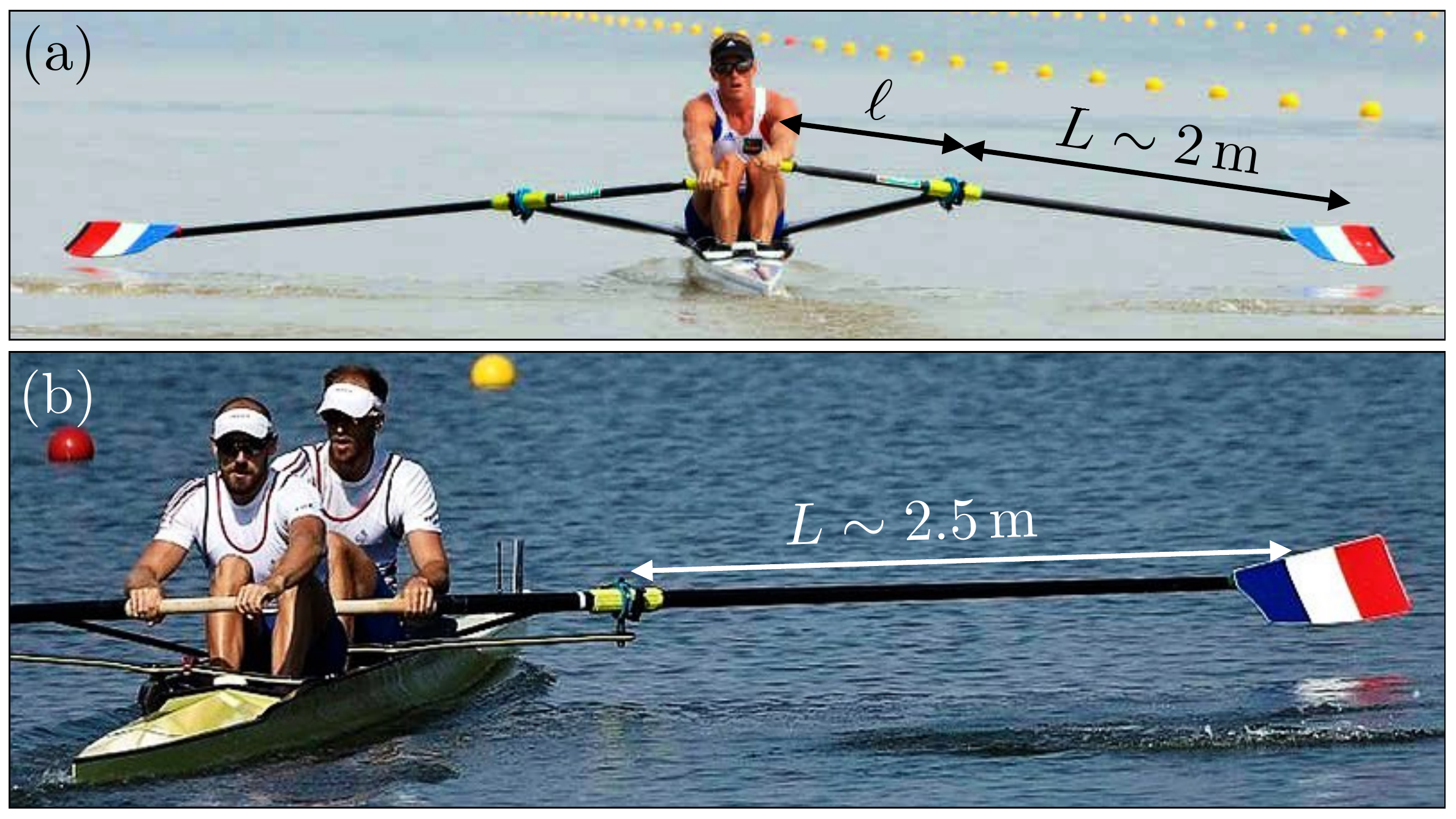}}
\caption{Front view of a single scull (a) and a coxless pair (b) \cite{nolte2005rowing, mcarthur1997high}. The inboard and outboard lengths are respectively denoted $\ell$ and $L$. Typical lengths are indicated.
}
\label{Fig1}
\end{figure}

\begin{figure}[t]
\centerline{\includegraphics[width=0.6\columnwidth]{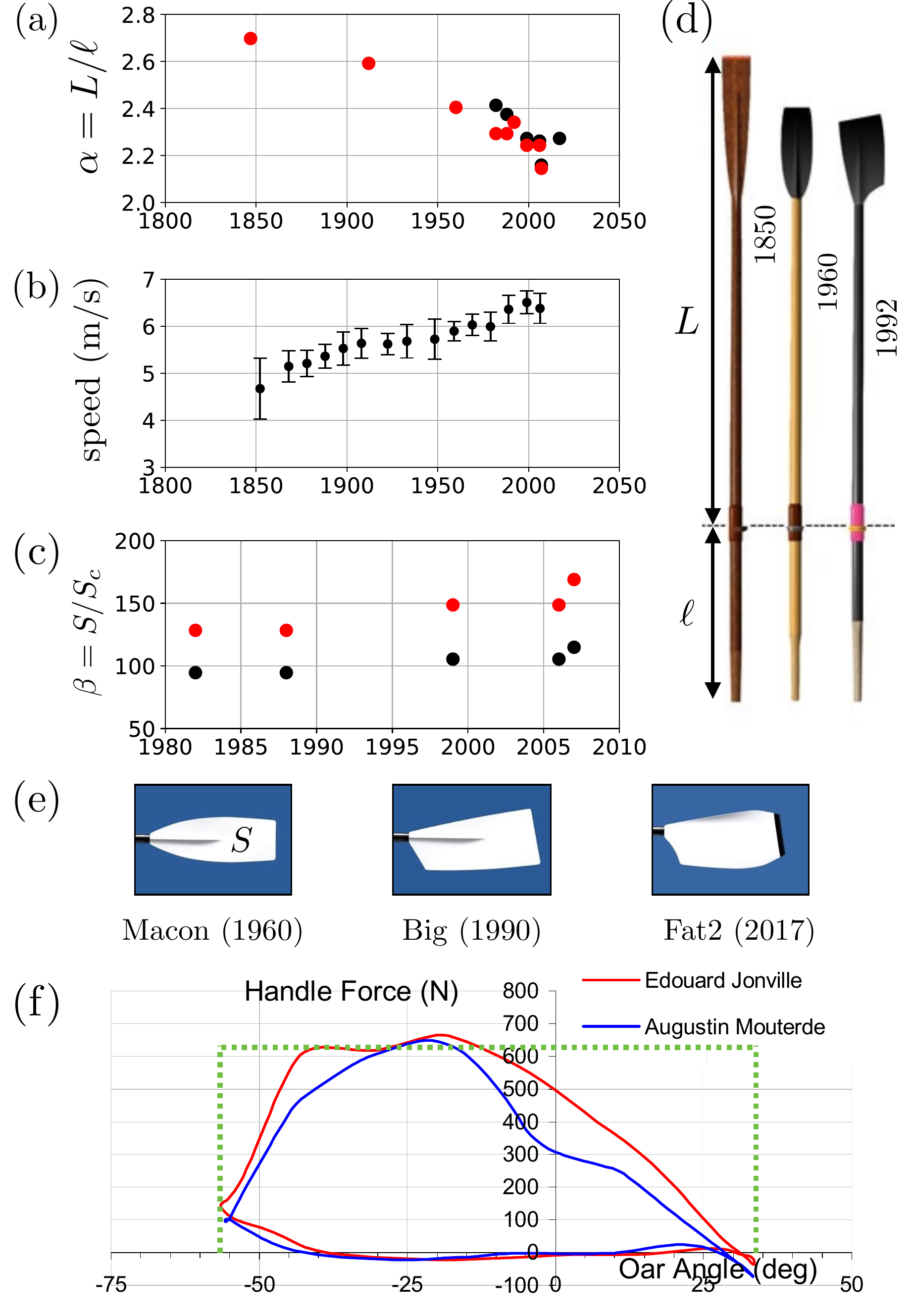}}
\caption{ (a) Evolution of the row aspect ratio $\alpha= L/\ell$ for sculling oars in black and sweep oars in red. Note that since the inboard length $\ell$ remained quite constant through time, $\alpha$ can be seen as the dimensionless row length. The oldest data points were obtained from race photographs, while the more recent ones come from \cite{nolte2009shorter} or were provided by the French athlete Thomas Baroukh. 
{(b) Evolution of the mean speed of the winner boat at the Oxford and Cambridge Boat Race (data gathered from \cite{wikipedia_boat_race}).}
(c) Evolution of the dimensionless blade area $\beta = S/S_c$ with $S$ the blade area and $S_c = S_\textrm{h}C_\textrm{h}/(N C_\textrm{d})$, where N is the number of blades, $C_\textrm{d}$ the drag coefficient of a blade, $S_\textrm{h}$ the hull wetted surface and $C_\textrm{h}$ the hull drag coefficient (for all the points, the hull wetted surface is taken constant as that of a coxless four rowing boat $S_h = 5.92$ m$^2$). The black dots are for sculling blades and red dots for sweep blades. Data come from \cite{nolte2009shorter}.
(d)  Picture of three different sweep oars (taken from \cite{wikipedia_rame}). The first oar dates back to 1850, the second one to 1960, and the last one to 1992. 
(e) Pictures, from left to right, of a \emph{Macon Blade} (1960), a \emph{Big Blade} (1990) and a \emph{Fat2 Blade} (2017) (from \cite{concept2}). 
(f) Handle force during one stroke as a function of the oar angle for two top-level French rowers (Edouard Jonville and Augustin Mouterde). These data were collected by Valery Klesnev \cite{biorow}. A $0^{\circ}$ oar angle corresponds to oars perpendicular to the boat. 
}
\label{evol}
\end{figure}

Most sports require different equipment for different weight categories and genders. For example, in shot put, women use masses of 4~kg, while men use masses of 7.5~kg. However in rowing sports \cite{nolte2005rowing, mcarthur1997high}, row characteristics are surprisingly quite constant in each discipline regardless of athletes strength and gender. In \emph{sculling} (Fig.~\ref{Fig1}(a)), the row size ranges from $287$ to $291$ cm \cite{nilson1987fisa,concept2}. For sweep boats (Fig.~\ref{Fig1}(b)), the row size reads $371$ to $376$\,cm \cite{nilson1987fisa,concept2}. 
Through rowing history, the tendency has been to reduce row lengths (by almost 25$\%$ since 1850, see Figs.~\ref{evol}(a) and \ref{evol}(d)). This evolution is also related to an increase in the blade area and the shift to asymmetric blades (Figs.~\ref{evol}(c) and \ref{evol}(e)).

In rowing competitions, the average stroke rate ranges between 30 and 40 strokes per minute depending on the boat category \footnote{The average stroke rate is about 33-35 strokes per minute for a single scull and 39 for a coxless four.}, which corresponds to strokes of 1.5 to 2 seconds. At the beginning of the race, the stroke rate is yet much higher (40-45 strokes per minute for a single scull and 45-50 for a coxless four) \cite{fede_int}. 
The rowing stroke is divided into two phases: a propulsive phase of about 0.7 seconds (40\% of the stroke) and a recovery phase of 1.1 seconds (60\% of the stroke) \cite{kleshnev2016biomechanics}.
During the propulsive stroke, typical force profiles exerted by the blade on the water were measured by Valery Klesnev \cite{biorow} and are reprinted in figure \ref{evol}(f). As one can see, the maximal handle force exerted is around 700~N. 
More interestingly, as described by coaches, a good rowing stroke corresponds to a force profile as constant as possible. 
Our study is conducted in this limit.  


Volker Nolte \cite{nolte2009shorter} performed an empirical study of the effects of row length on a dataset of rowing races. He reported that 'Shorter Oars Are More Effective'.
However, Laschowski \emph{et al.} \cite{laschowski2015effects} studied experimentally the effect of oar-shaft stiffness and length with elite athletes.
They showed that changes in stiffness and length of the oar led to small differences in the measured boat acceleration but these differences remained of the same order of magnitude as inter-stroke fluctuations.
Caplan \emph{et al.} \cite{caplan2007mathematical}, Leroyer \emph{et al.} \cite{leroyer2008experimental}, and H\'emon \cite{hemonhydrodynamic} took interest in the effects of row blade shapes by comparing real oars and highlighted the complexity of addressing such a problem. A number of authors \cite{ baudouin2002biomechanical, baudouin2004investigation, smith2002biomechanics, colloud2006fixed, sprague2007force, zatsiorsky1991mechanics, elliott2002rowing} addressed the problem of rowing efficiency and optimal rowing movement from the biomechanical perspective. 
In particular, Kleshnev \emph{et al.} \cite{kleshnev1999propulsive, kleshnev2016biomechanics} performed an intensive experimental study on propulsive efficiency varying oar travel, handle force, stroke rate and many other parameters.\\

Here we present a minimal and self-consistent analysis of the effects of row length and blade size on rowing performance, with a particular focus on rowing (fixed rowlock).  
{ We decouple the physics from its physiological counterpart and address the problem with imposed force instead of imposed kinematics.  
We propose a simple theoretical model that is compared to experiments made on a dedicated rowing robot.} 
In section 1, we present our rowing robot and our experimental results. In section 2, we derive the dynamical equations for a rigid row. In section 3, we derive the row velocity as function of imposed force for a single row attached to a static boat. In section 4, we compute the boat velocity at given imposed force for varying row lengths and compare our results to the experiments. In section 5, we present master plots on the efficiency of rowing boats and discuss the particular case of sweep oars optimisation.


\section{Robot rowing boat at constant force}
\label{Exp_boat}

\begin{figure}[b]
\centerline{\includegraphics[width=0.6\textwidth]{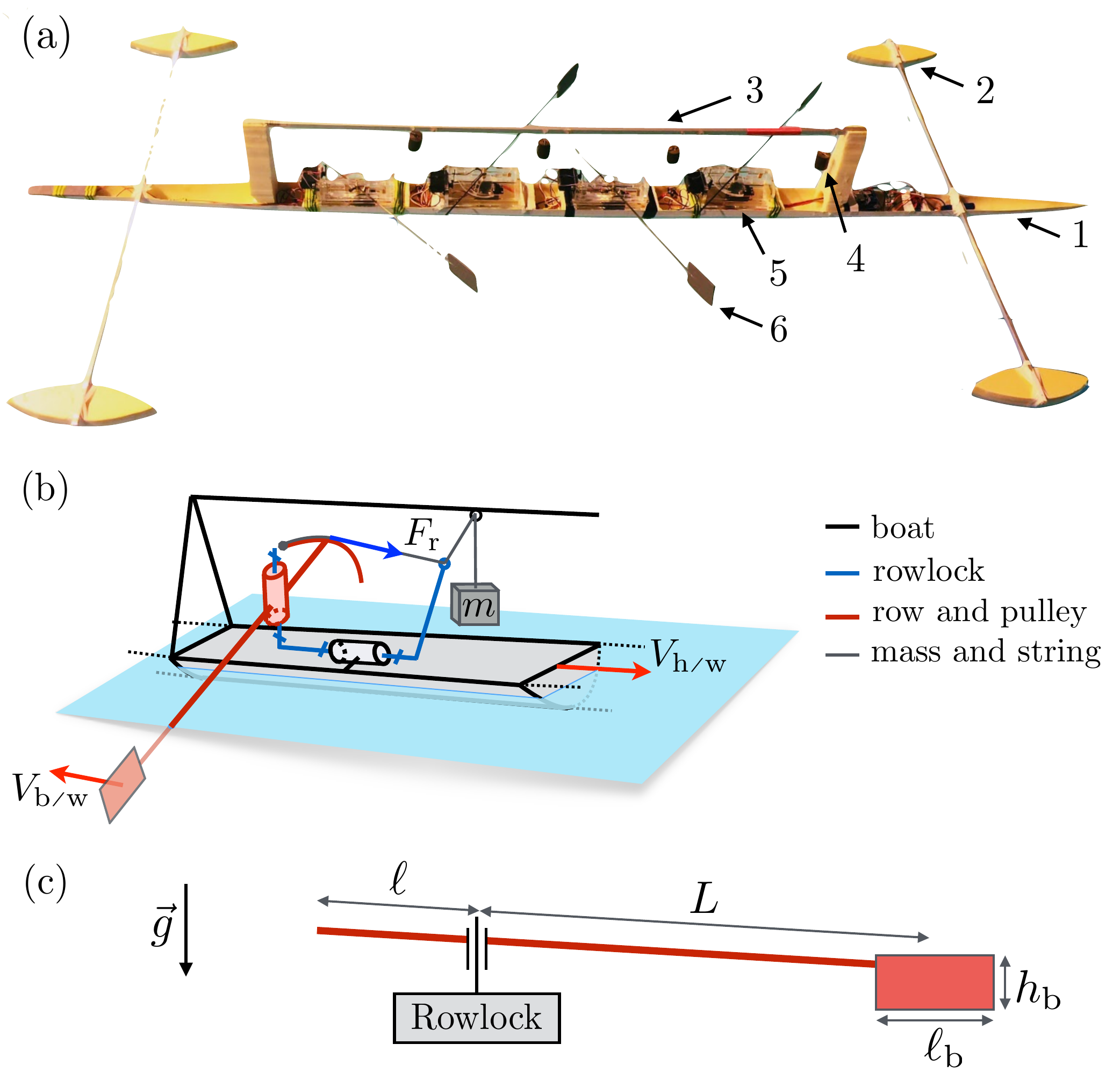}}
\caption{(a) Picture of the 2\,m long model rowing boat with 4 robot rowers at constant force with (1) a hull, (2) 4 floats, (3) a mass support, (4) 4 masses, (5) 4 robot rowers and (6) 4 rows. (b) Sketch of the mechanism of one robot rower. The row and pulley (red) rotate with respect to the rowlock (blue), itself in rotation with respect to the hull (black) to ensure lifting/dropping of the row between the propulsive and recovery phases.  A suspended mass/string system (gray) ensures row motion at constant force during the propulsive phase. The recovery phase and the blade flips were ensured by two servomotors and position sensors connected to an {Arduino}{\texttrademark} board (not shown for clarity). (c) Side view sketch of the row/rowlock system.}
\label{bateau}
\end{figure}

In this section, we introduce our model experiment, together with the corresponding experimental results. 
In order to understand the effect of the ratio $\alpha=L/\ell$ on the boat speed in the limit of constant force, we designed and manufactured a robot rowing boat with imposed propulsive force (see Fig.~\ref{bateau}). Using a homemade wooden mold based on a real rowing shell  \cite{vespoli1995eight} at the scale $1/10^\textrm{\scriptsize th}$, we built a glass fiber  rowing boat (see Fig.~\ref{bateau}(a)(1)) with 4 robot rowers (Fig.~\ref{bateau}(a)(5)) with one oar each (Fig.~\ref{bateau}(a)(6)). Constant force during the propulsive phase was ensured through a pulley-mass system. Each row was linked to a pulley centered at its rowlock. A suspended mass $m=80$\,g  (see Fig.~\ref{bateau}(a)(4)) was connected to the pulley through a string (see Fig.~\ref{bateau}(b)) by that setting the row in motion at constant force $F_\textrm{r}=mg$ (if we neglect frictional losses in all connections). The angular travel of the row was fixed to $\theta_0=90^\circ$. The recovery phase and the blade flips were ensured by two servomotors and position sensors connected to an {Arduino}{\texttrademark} board. The masses are suspended to a unique support (see Fig.~\ref{bateau}(a)(3)) and four polystyrene floats (see Fig.~\ref{bateau}(a)(2)) were added to ensure stability of the boat.\\

\begin{figure}[t]
\centerline{\includegraphics[width=0.6\textwidth]{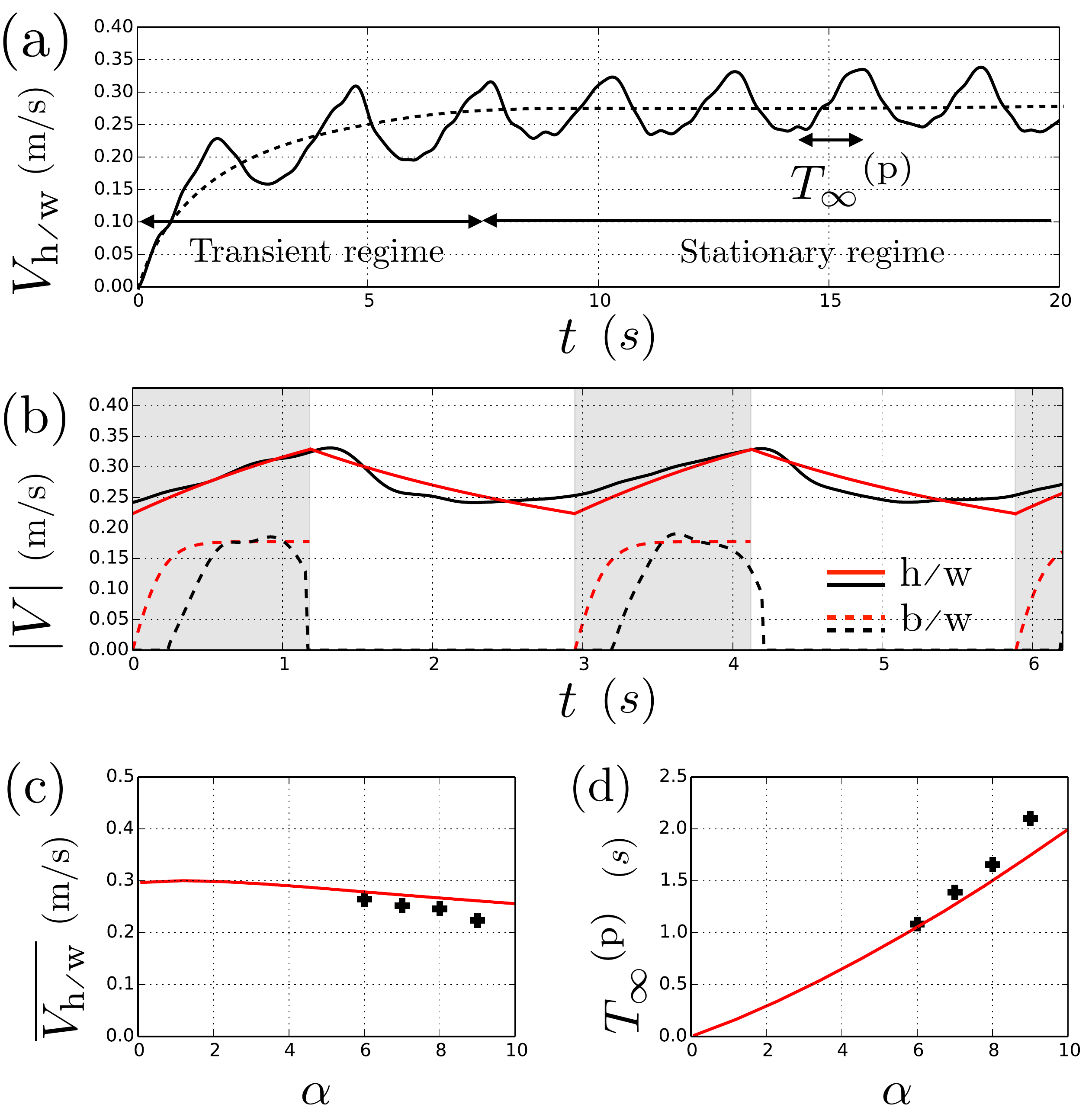}}
\caption{(a) Instantaneous velocity (solid lines) and mean velocity (dash line) of the hull with respect to the water, starting from rest at $t=0$, for $\alpha = 7$. 
(b) Velocity of the hull with respect to the water (solid lines) and absolute velocity of the blade (dash line) with respect to the water as function of time over two consecutive rowing cycles in the stationary regime, for $\alpha = 7$ (theory in red and experiments in black). (c) Mean boat velocity as function of $\alpha$. (d) Propulsive stroke duration as function of $\alpha$.
}
\label{dynamic}
\end{figure}

The experiments were performed at the Ecole polytechnique swimming pool. Setting the recovery phase time to a constant value $T^\textrm{(r)}=1.3$\,s, we video recorded the model boat rowing over a 25 m distance for four different row lengths, with corresponding aspect ratios  spanning from $\alpha=5$ to $\alpha=8$. As one can see in Fig.~\ref{dynamic}(a) after the start, the speed of the hull increases for about 8 seconds (3 strokes) until it reaches a stationary regime where the average speed is constant.	
Stationary stroke duration ${T_\infty}^{\hspace{-1mm}\textrm{(p)}}$ was recorded for each stroke using the position sensors mentioned above and averaged for each race.\\

In Fig.~\ref{dynamic}(b), the evolution of the instantaneous velocity with time  in the stationary regime is plotted (black curve). One can easily distinguish the two phases: the propulsive phase where the speed increases and the recovery phase where the speed decreases. {Note that in reality, the speed keeps increasing at the beginning of the recovery stroke due to the motion of the rowers on the boat \cite{boucher2017row}.} 
The experiments show that, when increasing $\alpha$, the average hull velocity $\overline{V_\textrm{h\sh w}}$ decreases (black dots in Fig.~\ref{dynamic}(c)), coherent with an increase in the propulsive stroke duration ${T_\infty}^{\hspace{-1mm}\textrm{(p)}}$ (Fig.~\ref{dynamic}(d)).
This observation agrees quite well with the historical evolution of the ratio $\alpha$ for real oars, presented in Fig.~\ref{evol}(a), as $\alpha$ decreased over the years with faster and faster boats (see Fig.~\ref{evol}(b)).\\

In the next parts, we derive a theoretical model to understand these results and predict the optimal row characteristics.

\section{Dynamics of a rigid oar}
\label{dynamics_rigid_oar}

\begin{figure}[b]
\centerline{\includegraphics[width=0.6\columnwidth]{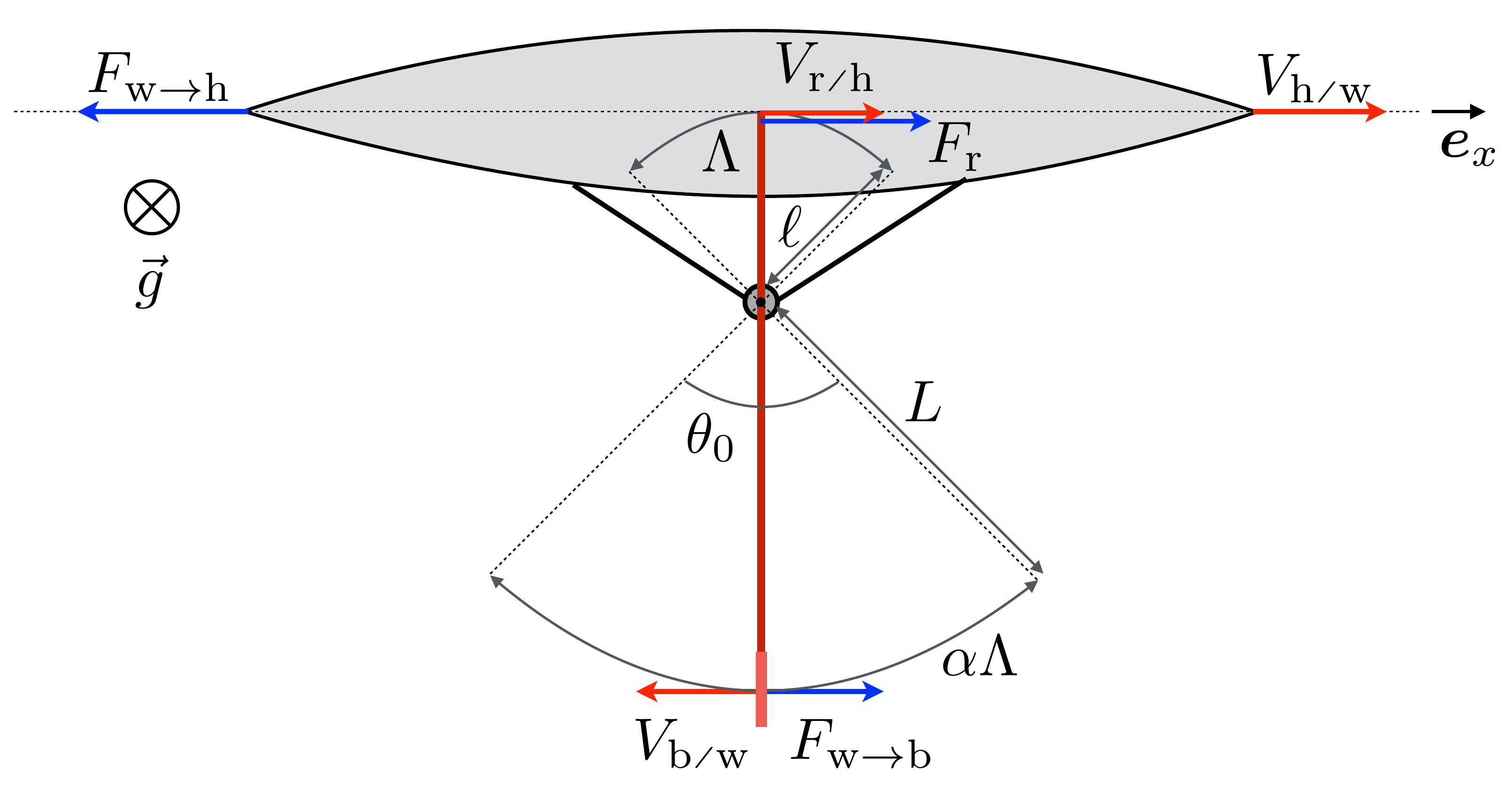}}
\caption{Top view sketch of a model rowing boat (for clarity, only one row/rowlock system is depicted). Forces are depicted in blue and velocities in red. }
\label{sketch}
\end{figure}

Here, we present the dynamical equations that govern oar propulsion for a given force profile exerted by the rower. The first kinematic relation relating the velocities in the different reference frames reads (see Fig.~\ref{sketch}):
\begin{eqnarray}
\boldsymbol{V}_\textrm{b\sh w}&=& \boldsymbol{V}_\textrm{b\sh h}+\boldsymbol{V}_\textrm{h\sh w} \ ,
\end{eqnarray}
where  $\boldsymbol{V}_\textrm{b\sh w}$, $\boldsymbol{V}_\textrm{b \sh h}$ and $\boldsymbol{V}_\textrm{h\sh w}$ respectively denote the speed of the blade with respect to the water,  the speed of the blade with respect to the hull, and the speed of the hull with respect to the water \footnote{The velocity of the blade is the velocity at the center point of the blade (where the hydrodynamic force is exerted).}. The second kinematic relation ensures conservation of angular momentum of the row at the oarlock (see Fig.~\ref{sketch}):
\begin{eqnarray}
\boldsymbol{V}_\textrm{b\sh h} = -{\alpha}\boldsymbol{V}_\textrm{r\sh h} \ ,
\end{eqnarray}
where $\boldsymbol{V}_\textrm{r\sh h}$ denotes the speed of the rower hands in the reference frame of the hull. 
{From now on, we assume that $\boldsymbol{V}_\textrm{b\sh w}$, $\boldsymbol{V}_\textrm{b \sh h}$, $\boldsymbol{V}_\textrm{h\sh w}$ and $\boldsymbol{V}_\textrm{r\sh h}$ are all parallel to the direction of motion of the hull, so we write: $\boldsymbol{V}_\textrm{b\sh w} = V_\textrm{b\sh w} \boldsymbol{e}_x$, $\boldsymbol{V}_\textrm{b \sh h} = V_\textrm{b \sh h} \boldsymbol{e}_x$, $\boldsymbol{V}_\textrm{h\sh w} = V_\textrm{h\sh w} \boldsymbol{e}_x$ and $\boldsymbol{V}_\textrm{r\sh h} = V_\textrm{r\sh h} \boldsymbol{e}_x$, with $\boldsymbol{e}_x$ the unit vector in the direction of motion of the boat.}
The forces exerted on the moving blade are ($i$) the pressure drag $F_\textrm{p}$, and ($ii$) the added mass $F_\textrm{am}$, both parallel to the blade motion in the reference frame of the water:
\begin{subeqnarray}
F_\textrm{p} &=&- \frac12 \rho S C_\textrm{d}|V_\textrm{b\sh w}|V_\textrm{b\sh w} \slabel{} \\
F_\textrm{am} &=& - \rho  \Omega  C_\textrm{m} \dot V_\textrm{b\sh w} \ ,\slabel{} \label{forces}
\end{subeqnarray}
where $\rho$ denotes the water density, $S = \ell_\textrm{b} h_\textrm{b} $ is the surface of the blade, $C_\textrm{d}$ and $C_\textrm{m}$ are the drag and added mass coefficients, and $\Omega = \pi S \ell_\textrm{b}/4$ is the volume of the cylinder with diameter $\ell_\textrm{b}$ and height $h_\textrm{b}$ (Fig.~\ref{bateau}(c)). {Note that we here neglect all contributions related to lift forces on the blade \footnote{Although lift might not be negligible especially during the beginning and the end of the rowing stroke \cite{caplan2007fluid,baudouin2002biomechanical}, it has the same scaling as the drag force and thus taking it into account would not significantly change our results.}.}
The net force $F_{\textrm{w} \rightarrow \textrm{b}} = F_\textrm{p} + F_\textrm{am}$ exerted by the water on the blade must match that of the rower  $F_\textrm{r} $ through a torque conservation relation at the rowlock (assuming the oar tubes to be rigid and of negligible mass). 
That is:
\begin{eqnarray}
 F_{\textrm{w} \rightarrow \textrm{b}} &=&  \frac{1}{\alpha} F_\textrm{r} \ . \label{torque_alpha}
\end{eqnarray}
We then nondimensionalise the problem by letting $V = \hat V V_\textrm{c}$, $t = \hat t \tau_\textrm{c}$ and $F=\tilde FF_\textrm{c}$, where $V_c = \sqrt{{2  F_\textrm{c}}/(\rho S C_\textrm{d})} $, $\tau_\textrm{c} =C_\textrm{m}\Omega  \sqrt{{2 \rho }/(F_\textrm{c} S C_\textrm{d})}$, and $F_{\textrm c}$ is  a characteristic force scale. The natural characteristic length of the problem $L_\textrm{c} = {V_\textrm{c} \tau_\textrm{c}} = 2 C_{\textrm m} \Omega/(SC_\textrm{d})$ compares the effects of added mass and pressure drag. Using Eqs.~(\ref{forces}) and (\ref{torque_alpha}), one obtains:
\begin{eqnarray}
\label{dyn_eq}
|\hat{V}_\textrm{b\sh w}|\hat{V}_\textrm{b\sh w} +\dot{\hat{V}}_\textrm{b\sh w} = - \frac{1}{\alpha}\tilde{F}_\textrm{r} \ .
\label{eq_stat}
\end{eqnarray}

Equation~(\ref{dyn_eq}) can be solved numerically for any force profile $\tilde{F}_\textrm{r}(t)$, such that one can determine the exact blade velocity $\hat{V}_\textrm{b\sh w}$.

\section{Single oar dynamics at constant force} 
\label{staticsect}
In the following, we choose to work on a simple and analytically solvable case by assuming a constant imposed  force (Fig.~\ref{evol}(f)). Although previous studies (see \cite{baudouin2004investigation, jones2002mechanics, baudouin2002biomechanical}) show evidence of slightly time-dependent force profiles, we here wish to extract the general physics and scaling arguments of rowing mechanics with minimal ingredients, for which a constant force seems appropriate from a physiological point of view. 

Letting the deployed force of the rower $F_\textrm{r} =  F_\textrm{c}$, namely $\tilde F_\textrm{r}=1$ into Eq.~(\ref{dyn_eq}), together with $\hat{V}_{\textrm b\sh \textrm w}(0) = 0 $ yields: 
\begin{eqnarray}
\hat{V}_{\textrm b\sh \textrm w}(\hat{t} \hspace{0.03cm}) = -\frac1{\sqrt{\alpha}} \, \mathrm{tanh}\left(\frac{\hat{t}}{\sqrt{\alpha}}\right) \ .
\label{sol_v}
\end{eqnarray}
In particular, one has $\hat{V}_{\textrm b\sh \textrm w}(\hat t\ll 1) =-\hat t/{\alpha}$ and $\hat{V}_{\textrm b\sh \textrm w}(\hat t\gg 1) 
=-1/\sqrt{\alpha}$.\\

\begin{figure}[t]
\centerline{\includegraphics[width=0.6\textwidth]{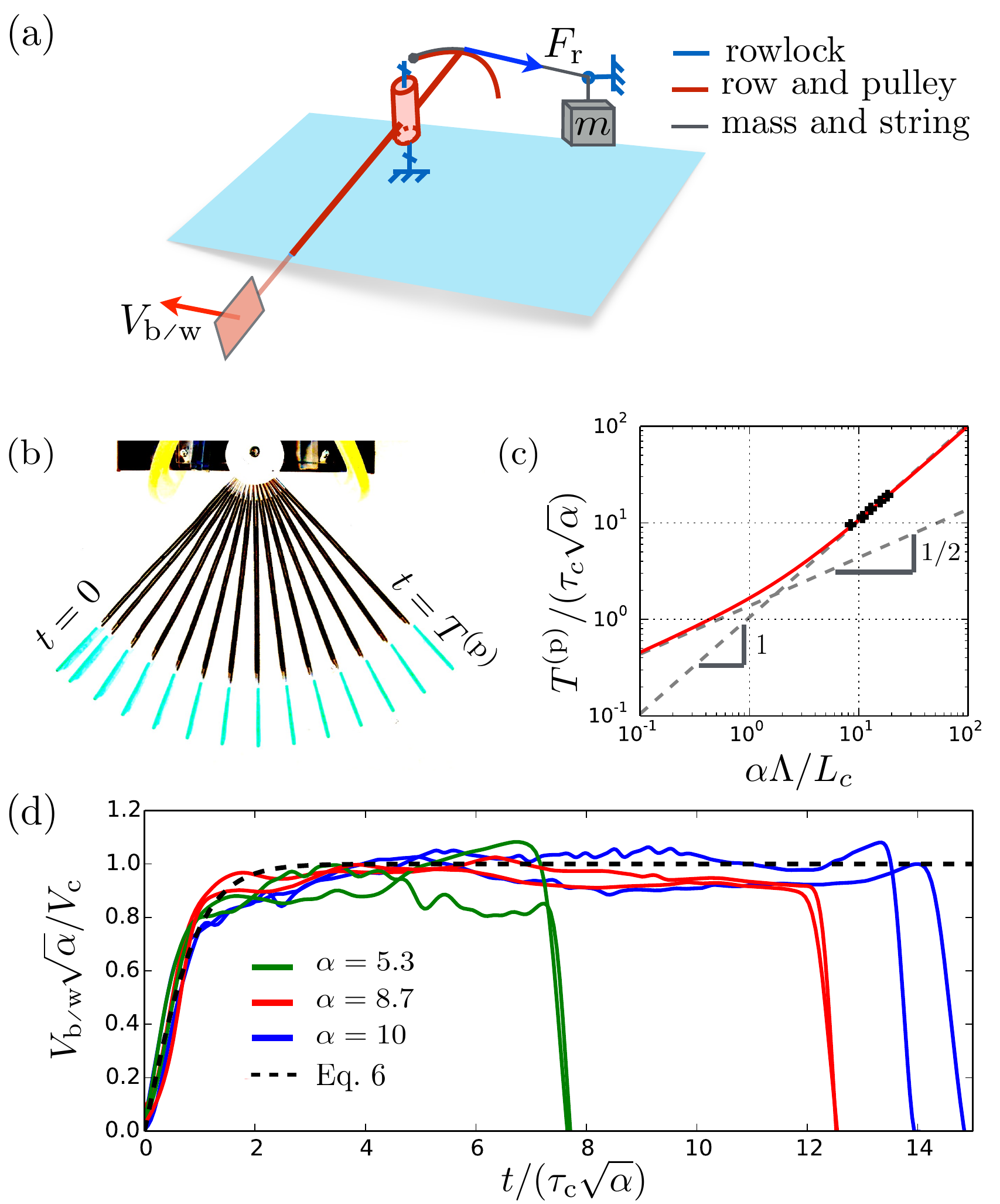}}
\caption{(a) Sketch of the mechanism for the static boat experiment. The row and pulley (red) rotate with respect to the rowlock (blue) which is fixed in the reference frame of the lab.  A suspended mass/string system (gray) ensures row motion at constant force. (b) Chronophotography of the static boat experiment with a 20 cm row (corresponding to $\alpha=5.3$). The time between each frame is $60$\,ms.  (c) Rescaled stroke duration as function of rescaled stroke length. The red curve corresponds to Eq.~(\ref{eq5}) and the black crosses signify experimental data with constant $F_0= 5$N while varying $\alpha$. (d) Dimensionless velocity as function of dimensionless time for three different values of $\alpha$, each with two values of $F_0\in\{5\rm{N},10\rm{N}\}$. }
\label{result_statique}
\end{figure}

In order to validate our model, we first consider that the rowlock is immobile in the reference frame of the water ($V_{\textrm h\sh \textrm w} = 0$). This situation corresponds to a fixed boat, a hypothesis that we shall relax in Sect.~\ref{ship_prop}. To check Eq.~(\ref{sol_v}) experimentally, we performed a simple experiment involving one row subjected to a constant force (Fig.~\ref{result_statique}(a)). The force was exerted by a reference mass $m$ suspended to a nylon string, itself connected to the row handle through a pulley (so that $F_\textrm{r}=mg$ where $g$ denotes the acceleration of gravity).  
The rowlock was attached to the basin boundary. A top view image is presented in Fig.~\ref{result_statique}(b). For a given mass $m$ we measured the velocity of the blade while varying $L = 5- 30$\,cm at constant $\ell = 3$\,cm (which amounts to varying $\alpha$). The blade dimensions were $\ell_{\rm b}  = 7.0$\,cm and $h_{\rm b} = 4.7$\,cm. Fitting  the theory (Eq.~(\ref{sol_v})) to the experimental results (Fig.~\ref{result_statique}(d)) led to $C_\textrm{d} = 2.0 \pm 0.2$ and $C_\textrm{m} = 0.7 \pm 0.1$ in good agreement with literature values (for a plate of ratio height to span around 0.6, one has $C_\textrm{d} \simeq 1.2$ \cite{hoerner1965fluid} and $C_\textrm{m} \simeq 0.70$ \cite{blevins1980formulas}). 
\\

To go one step further, we compute the stroke duration $T^\textrm{(p)}$. The travel of the row end (held by the rower hands) is given by   $\Lambda = \theta_0\ell$ where $\theta_0 = 90^\circ$ (Fig.~\ref{sketch}). The stroke duration $T^\textrm{(p)}$ solves:
\begin{eqnarray}
	\int_0^{T^\textrm{(p)}} V_{\textrm r\sh \textrm h} \mathrm{d}t  = \Lambda  \ . \label{period_row}
\end{eqnarray}
Note that in the setup working at constant force $ F_{\rm r}$ amounts to working at constant rower energy over a cycle $E_\textrm{r} = \Lambda F_{\rm r}$. Using Eq.~(\ref{sol_v}), one obtains:
\begin{equation}
{T^\textrm{(p)}} = \tau_\textrm{c}\sqrt{\alpha}\ \mathrm{cosh}^{-1}e^{  \alpha \Lambda / L_\textrm{c}} \ .
\label{eq5}
\end{equation}
Note that $\alpha \Lambda/ L_\textrm{c}$ is the dimensionless number that compares the travel of the blade $\alpha \Lambda$ and the characteristic length $L_\textrm{c}$. As such $L_\textrm{c}$ can be interpreted as the length above which the limit velocity  is reached and added mass no longer plays a role. Figure~\ref{result_statique}(c) displays the theoretical rescaled stroke duration as function of $\alpha \Lambda / L_\textrm{c}$ as well as the experimental data points. The stroke duration increases with $\alpha$, consistent with increasing blade travel $\alpha \Lambda$ and decreasing blade propulsive force $F_\textrm{r}/\alpha$. Decreasing $\alpha$ amounts to increasing rowing frequency.  Two regimes can be distinguished: an added mass dominated regime corresponding to $\alpha \Lambda /L_\textrm{c} \ll 1$ for which $T^\textrm{(p)}\sim \alpha$, and a pressure drag dominated phase for which $\alpha \Lambda /L_\textrm{c} \gg 1$ and $T^\textrm{(p)}\sim \alpha^{3/2}$ (note that the experimental data on Fig.~\ref{result_statique} (c) lies on the pressure drag dominated phase). 

\section{Boat propulsion at constant force}
\label{ship_prop}

In this section we relax the constraint of an immobile hull (${V}_{\textrm h\sh \textrm w} \neq 0$). The additional equation needed to close the problem results from the force balance on the hull. This is $ F_{\textrm{w} \rightarrow \textrm{h}} =  N F_\textrm{r}/\alpha $, with $N$ the number of blades.  
We assume that the drag force on the hull is dominated by skin friction \footnote{ The skin friction is expected to account for 80\% of the overall drag \cite{baudouin2002biomechanical,mcmahon1971rowing}. Thus, we neglect here the other contributions to the drag on the boat (form drag, wave drag and aerodynamic drag). 
} and we do not take into account the motion of the rowers on the boat.
According to Newton's second law, one obtains in this limit:
\begin{eqnarray}
M \dot{V}_\textrm{h\sh w} + 1/2 \rho S_\textrm{h} C_\textrm{h} |V_\textrm{h\sh w}|V_\textrm{h\sh w} =  N F_\textrm{r}/\alpha \ ,
\end{eqnarray}
where $M$ is the mass of the boat, $S_\textrm{h}$ the wetted surface of the hull and $C_\textrm{h}$ its skin drag coefficient.
In a different way than in Sect.~\ref{dynamics_rigid_oar} -- as we want to assess the effect of oar parameters for a given boat -- we now use the hull parameters to non-dimensionalise the problem. Thus, we introduce a new velocity scale $V^\star = \sqrt{{2  NF_\textrm{c}}/(\rho S_\textrm{h}C_\textrm{h})} $ and a new time scale $\tau^\star =M  \sqrt{{2}/( \rho NF_\textrm{c} S_\textrm{h}C_\textrm{h})}$ and we write $V = \tilde V V^\star$, $t = \tilde t \tau^\star$ and $F=\tilde FF_\textrm{c}$, 
with $F_\textrm{c}$ the characteristic force introduced in Sect.~\ref{dynamics_rigid_oar}. The natural characteristic length of the problem is now $L^\star = {V^\star \tau^\star}$. The dimensionless equation governing the boat velocity then writes:

\begin{eqnarray}
\label{dyn_eq_h}
 |\tilde{V}_\textrm{h\sh w}|\tilde{V}_\textrm{h\sh w} + \dot{\tilde{V}}_\textrm{h\sh w} = \frac{1}{\alpha} \tilde{F}_\textrm{r} \ .
\end{eqnarray}


With the new set of characteristic parameters, the dimensionless equation governing the dynamics of the oar (Eq.~(\ref{eq_stat})) must accordingly be changed to:
\begin{eqnarray}
\beta |\tilde{V}_\textrm{b\sh w}|\tilde{V}_\textrm{b\sh w} +\gamma \dot{\tilde{V}}_\textrm{b\sh w} = -\frac{1}{\alpha} \tilde{F}_\textrm{r} \ ,
\end{eqnarray}
where:
\begin{eqnarray}
\beta = N S C_\textrm{d}/(S_\textrm{h}C_\textrm{h})
\label{eq_beta}
\end{eqnarray}
denotes the ratio between the blades' pressure drag and the hull skin drag and $\gamma =  N \rho  \Omega  C_\textrm{m} /M$ is the ratio between the blades' added mass and the boat mass. In the following and for the sake of simplicity, we consider self-similar blades (ratio $h_\textrm{b}/\ell_\textrm{b}$ constant), so that $\gamma \sim \beta^{3/2}$, by that reducing the number of dimensionless parameters.\\

Each rowing cycle $k$ is made of two phases: ($i$) the propulsive phase at constant force with duration ${T_k}^{\hspace{-1mm}(\textrm{p})}$ for which we set  $\tilde{F}_\textrm{r}=1$ and ($ii$) the recovery phase with duration ${T_k}^{\hspace{-1mm}(\textrm{r})}$ for which $\tilde{F}_\textrm{r}=0$. 
The overall cycle period reads $ T_k = {T_k}^{\hspace{-1mm}(\textrm{p})}+{T_k}^{\hspace{-1mm}(\textrm{r})}$. In the following, we shall restrict to a constant and prescribed duration for the recovery phase ${T_k}^{\hspace{-1mm}(\textrm{r})}=T^{(\textrm{r})}$ 
\footnote{ Note that another possible choice would be to set ${T_k}^{\hspace{-1mm}(\textrm{r})}~=~{T_k}^{\hspace{-1mm}(\textrm{p})}$.}.
 The solution of Eq.~(\ref{dyn_eq_h}) reads in the propulsive phase  ($ \tilde t\in [\tilde t_k, \tilde t_k+  {\tilde{T}_k}^{\hspace{-0mm}(\textrm{p})}]$  with $\tilde t_k = k\tilde{T_k}$):  $\tilde{V}_{\textrm h\sh \textrm w}^{(\textrm{p})}(\tilde t\hspace{0.03cm}) =$
\begin{equation}
\label{}
 \frac1{\sqrt{\alpha}} \, \mathrm{tanh}\left[\textstyle {\frac{1}{\sqrt{\alpha}}}\, {(\tilde t-\tilde t_k)}  + \mathrm{tanh}^{-1 } \big[  \tilde{V}_{\textrm h\sh \textrm w}^{(\textrm{p})}(\tilde t_k) \sqrt{\alpha}\, \big] \right]  ,
\end{equation}
and in the recovery phase ($\tilde t\in[\tilde t_k+ {\tilde{T}_k}^{\hspace{-0mm}(\textrm{p})},\tilde t_{k+1}]$):
\begin{equation}
\tilde{V}_{\textrm h\sh \textrm w}^{(\textrm{r})}(\tilde t\hspace{0.03cm}) = \frac{1}{\left({\tilde{V}_{\textrm h\sh \textrm w}^{(\textrm{p})}\left(\tilde t_k+\tilde T^{(\textrm{p})}_k \hspace{0.03cm}\right)} \right)^{-1}+ { \left(\tilde t-\tilde t_k-{\tilde{T}_k}^{\hspace{-0mm}(\textrm{p})}\right)}  }  . \label{sol_hull}
\end{equation}
To close the system, one needs the continuity equation for the velocity:
\begin{equation}
\tilde{V}_{\textrm h\sh \textrm w}^{(\textrm{r})}(\tilde t_{k+1}) =  \tilde{V}_{\textrm h\sh \textrm w}^{(\textrm{p})}(\tilde t_{k+1}) \ ,
\end{equation}
and the equation for the stroke duration ${T_k}^{\hspace{-1mm}(\textrm{p})}$  of the $kth$ propulsive phase:
 \begin{equation}
 	\int_{t_k}^{t_k+T^{(\textrm{p})}_k} V_{\textrm b\sh \textrm h} \textrm dt  =\int_{t_k}^{t_k+T^{(\textrm{p})}_k} \left(V_{\textrm b\sh \textrm w} - V_{\textrm h\sh \textrm w}\right) \textrm dt =  -\alpha  \Lambda  \ .
\end{equation}

In order to test our theory, we compare its predictions with the results for our robot rowing boat with imposed propulsive force presented in the Sect. \ref{Exp_boat} (see Fig.~\ref{bateau}). The experimental results are reported in Fig.~\ref{dynamic} and compared to the theoretical predictions of our model. The row parameters $C_\textrm{d}$ and $C_\textrm{m}$ were  estimated in the previous section. The drag coefficient on the hull $C_\textrm{h}$ was estimated by measuring the deceleration of the fully loaded model boat with a given initial velocity and blades out of the water (we found $ S_\textrm{h}C_\textrm{h}=(2.2 \pm 0.1)\,10^{-3}$\,m$^2$). \\

The measured instantaneous hull velocity (Fig.~\ref{dynamic}(b)) is found in quite good agreement with the theoretical predictions. The  stroke duration (Fig.~\ref{dynamic}(d)) and the mean velocity (Fig.~\ref{dynamic}(c)) are slightly off the theoretical curves. These small discrepancies can be the results of two different effects. First, our model does not account for the dynamic  inclination of $F_\textrm{b\sh w}$ with respect to the direction of motion of the boat, by that overestimating the propulsive force. Indeed the instantaneous real propulsive force should read $F_\textrm{b\sh w}\cos\theta$ where $\theta\in [-\theta_0/2,\theta_0/2]$ denotes the angle of the row with respect to the normal to the direction of motion. Although we do not wish to increase the model's complexity further by accounting for this effect, the associated correction can be roughly estimated by $\langle \cos \theta \rangle_{[-\theta_0/2,\theta_0/2]}\approx 10\%$. Second, our robot rowing boat suffered from an abrupt slow down at the end of the propulsive phase (Fig.~\ref{dynamic}(b)) due to both ($i$) the rows hitting the mechanical stop before being lifted out of the water, and ($ii$) the deceleration of the masses increasing the drag on the hull. 
 Note that, in contrast with the static boat experiments of Sect.~\ref{staticsect}, our model boat lies on the crossover between the added mass and pressure drag dominated regimes. Indeed, as can be seen on Fig.~\ref{dynamic}(b), the blade velocities display roughly balanced acceleration and plateau timescales. This key effect is precisely due to relaxing the static constraint by that shortening the blade's travel with respect to the water.\\
 
To interpret the results in term of efficiency, we define the \emph{anchoring} $\mathcal A$ of the blade, as the ratio of the distance travelled by the hull during the propulsive phase, denoted $\Lambda_\textrm{h}(\alpha)$, and the travel of the blade in the reference frame of the boat $ \alpha \Lambda$ (see Fig.~\ref{sketch}): 
\begin{eqnarray}
\mathcal A &=&  \frac{\Lambda_\textrm{h}}{ \alpha \Lambda} \ , \ \textrm{  with } \ \Lambda_\textrm{h}(\alpha)={\int_0^{{T}^{\textrm{(p)}}_\infty}  V_{\rm h \sh w} \textrm{d}t} \ .\quad \label{accdef}
\end{eqnarray} 
The anchoring can be seen as the oar efficiency \footnote{The anchoring can be related to the instant centre of rotation of the oar.}. Indeed, if $\mathcal A=1$, the blade does not move with respect to the water and all the rower's energy is transferred to the boat. In contrast, if $\mathcal A=0$ the boat does not move and the oars slip in the water.
Interestingly the \emph{anchoring} has an energetic interpretation. The propulsive energy provided by the rower $E_\textrm{r} = \Lambda F_{\rm r}$ is dissipated by both the hull $E_\textrm{h}=\Lambda_{\rm h} F_{\rm r}/\alpha$ and the blades $E_\textrm{b}$, such that $E_\textrm{r}=E_\textrm{h}+E_\textrm{b}$.  Equation~(\ref{accdef}) yields: 
\begin{eqnarray}
\mathcal A &=&  \frac{E_\textrm{h}}{ E_{\rm r}} \ ,
\end{eqnarray}
that is: the anchoring $\mathcal A \in[0,1]$ quantifies the efficiency of energy transfer between the rower and the boat \cite{kleshnev1999propulsive,kleshnev2016biomechanics}. 


The tendencies and the optimisation are discussed in the following section and compared to real rowing boat data.


\section{Physical discussion}

\begin{figure}
\centerline{\includegraphics[width=0.4\columnwidth]{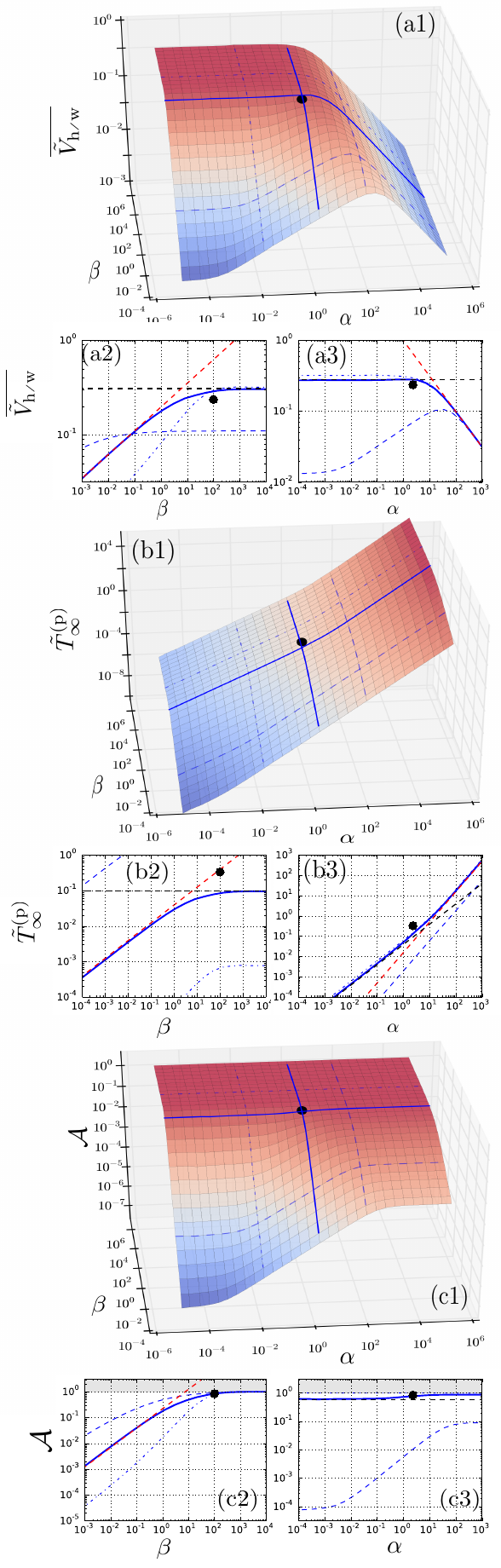}}
\caption{(a1-3) Rescaled mean boat velocity $\overline{\tilde{V}_{\textrm h  / \textrm w}}$, (b1-3) propulsive stroke duration ${\tilde{T}_\infty}^{(\mathrm{p})}$ and (c1-3) anchoring $\mathcal A$ as function of $\alpha$ and $\beta$. The manifolds (a1, b1, c1) and the curves obtained at given $\alpha$ (a2, b2, c2) or given $\beta$ (a3, b3, c3) were obtained numerically with the parameters of a coxless four rowing boat.  Asymptotes at small and large scale in (a2, a3, b2, b3, c2, c3) are indicated with black and red dashed lines. The black dot in each plot  corresponds to the observed data for a coxless four rowing boat.}
\label{figtorture}
\end{figure}

Here, we discuss the global optimisation problem as function of parameters $\alpha$ and $\beta$ and confront our results to real rowing boats. 
Figure~\ref{figtorture} displays the dimensionless hull velocity, the stroke duration and the anchoring as function of $\alpha$ and $\beta$ \footnote{Note that we here use the parameters of real rowing boats in order to be able to compare our theory to the empirical data.}, together with a few 2D cut to simplify the discussion. The velocity plot can be understood from the stroke duration and anchoring plots through the relation:

\begin{eqnarray}
V&\sim&  \frac{\mathcal A \alpha \Lambda }{T^{\textrm{(p)}}_\infty}\ .
\end{eqnarray}

 At constant $\alpha$, the rescaled velocity and the stroke duration are increasing functions of $\beta$ (Figs.~\ref{figtorture}(a2) and (b2)) and saturate at large $\beta$. This can be understood through the anchoring behavior (Fig.~\ref{figtorture}(c2)). At small $\beta$ -- small blades -- the anchoring is weak and much energy is dissipated by the blades motion with respect to the water. At large $\beta$, the large blades are well anchored in the water ensuring maximal energy transfer to the boat, or equivalently that the hull velocity matches the blade velocity with respect to the boat. 
 The behavior with $\alpha$ at constant $\beta$ is less trivial. The stroke duration is an increasing function of $\alpha$ and the velocity crosses over from a plateau at small $\alpha$ (added mass dominated) to an $\alpha^{-1/2}$ regime (pressure drag dominated) at large $\alpha$. At large given $\beta$ the anchoring is maximal ($\mathcal A \rightarrow 1$) and the velocity is a monotonous function of $\alpha$, while for  small given $\beta$ there exists an optimal value of $\alpha$ that maximises the velocity.\\
 
The mean power injected by a rower at constant maximal force writes:
\begin{eqnarray}
\bar{P} = \frac{1}{T^{\textrm{(p)}}_\infty } \int_0^{T^{\textrm{(p)}}_\infty }  {F}_\textrm{r} V_{\rm h \sh w} \textrm{d}t  = \frac{{F}_\textrm{r}\Lambda}{T^{\textrm{(p)}}_\infty}
\end{eqnarray}
$\bar{P}$ scales as $1/T_\infty^{(\rm{p})}$. Note that decreasing the dimensionless row length $\alpha$ decreases the stroke duration $T_\infty^{(\rm{p})}$ and thus increases the mean injected power. \\

On the one hand, if one wants to achieve maximum velocity regardless of injected energy -- or equivalently mean power -- (sprint strategy), one should choose rather short oars $\alpha \sim 1$  (at the limit of the plateau corresponding to the transition between the added mass and pressure drag dominated regimes (Fig.~\ref{figtorture}(a1)).
  However, bear in mind that short oars go hand in hand with high rowing frequency which might be hard to achieve from a physiological point of view, by that setting a lower bound to $\alpha$ \cite{hill1938heat, wilkie1949relation}. Furthermore, reducing the row length imposes to increase the blade surface to maximise the anchoring in water $\mathcal A$.
On the other hand, if one is rather tempted by maximal efficiency $\mathcal A \to 1$ (endurance race), then long oars are indicated in order to reduce the mean power provided by the rower (Fig.~\ref{figtorture}(c1)).\\

\begin{figure}[h]
\centerline{\includegraphics[width=0.6\columnwidth]{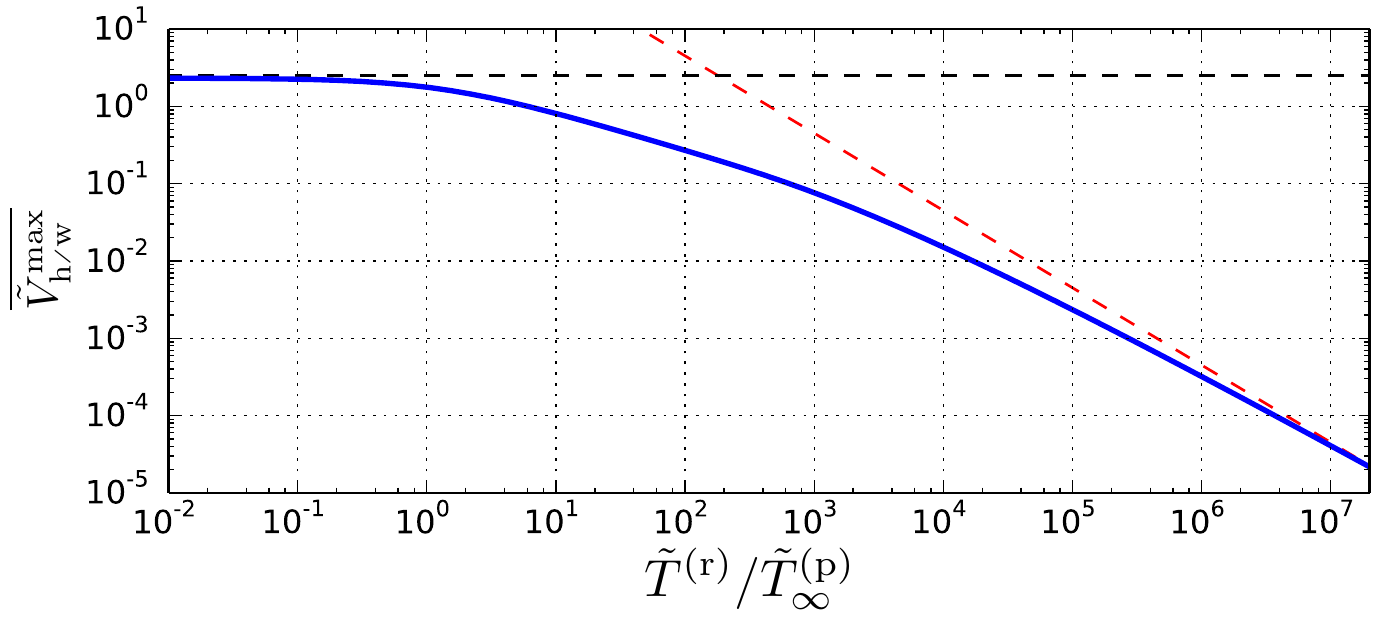}}
\caption{Rescaled maximal mean boat velocity (typically the mean velocity on the plateau, see Fig.~\ref{figtorture}(a1)) as function of recovery to propulsion time ratio.}
\label{temps_retours}
\end{figure}

Note that all the results presented in Fig.~\ref{figtorture} were obtained for a fixed recovery time $\tilde T^\textrm{(r)} = 1.1 $, roughly corresponding to that of a real rowing race. The recovery time plays a role on the position of the transition point between the different regimes, as well as on the maximal mean velocity $\overline{{\tilde{V}}^{\mathrm{max}}_\textrm{h\sh w}}$ reached in the plateau region. Figure~\ref{temps_retours} displays the maximal mean velocity as function of recovery to propulsion time ratio.
Decreasing the recovery time $\tilde T^\textrm{(r)}$ reduces the fluctuations of the boat velocity and leads to an increasing maximal mean velocity that saturates for $\tilde T^\textrm{(r)}/T^{\textrm{(p)}}_\infty \le 1$. 
In the limit $\tilde T^\textrm{(r)}/T^{\textrm{(p)}}_\infty \gg 1$,  the distance travelled by the boat during the propulsive phase reaches a constant value, while the period of the rowing cycle scales as $\tilde T^\textrm{(r)}$. Therefore, one has: $\overline{{\tilde{V}}^{\mathrm{max}}_\textrm{h\sh w}}\sim 1/\tilde T^\textrm{(r)}$.\\


\begin{figure}[b]
\centerline{\includegraphics[width=0.8\textwidth]{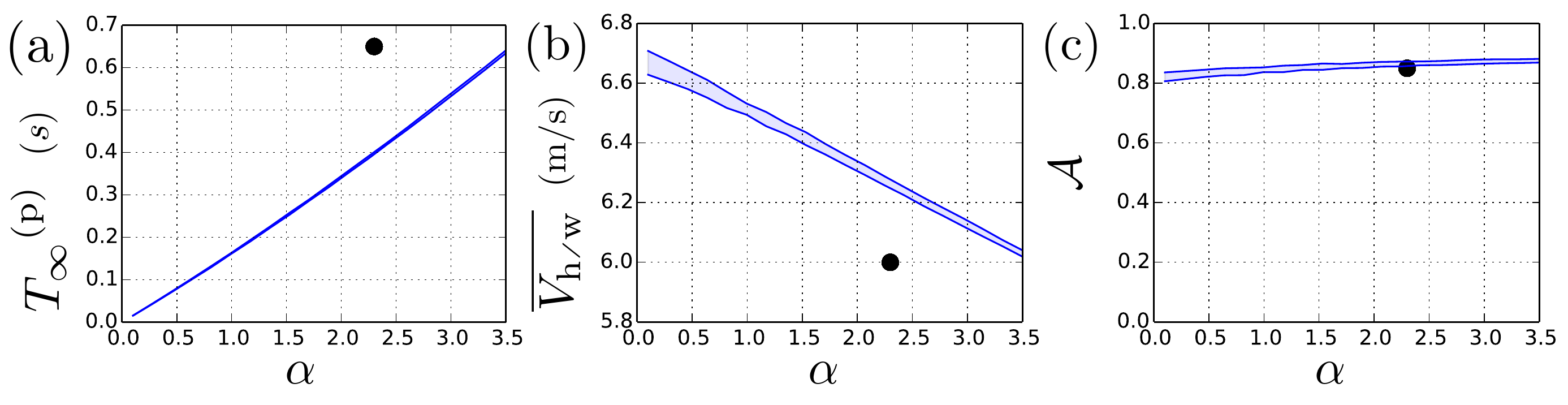}}
\caption{(a) Propulsive stroke duration as function of $\alpha$. (b) Mean boat velocity as function of $\alpha$. (c)  Anchoring (or equivalently energy transfer efficiency) as function of $\alpha$ (Eq.~(\ref{accdef})). The blue curves were obtained numerically for $\beta$ between $90$ and $120$. Empirical data for a coxless four rowing boat are indicated with a black dot. } 
\label{opt_reel}
\end{figure}

Now, let us focus on the case of a coxless four rowing boat. We consider that each rower deploys a force $F_r=700$\,N (Fig.\ref{evol}(f)).  The stroke duration, the mean boat velocity and the anchoring computed from our model in this specific case are presented in Fig.~\ref{opt_reel}. For real sweeping oars, $ \alpha \simeq  2.2$ and  $\beta \simeq 100 $ which lies precisely at the cross-over between the added mass and pressure drag regimes.
As one can see in Fig.~\ref{opt_reel}(c), the estimated anchoring for a coxless four rowing boat \cite{kleshnev1999propulsive} compares well with the theoretical anchoring predictions, with $\mathcal A$ being close to $80$ \%.

The real mean velocity (Fig.~\ref{opt_reel}(b)) is smaller than the theoretical one. This is due to all the assumptions of our model: in particular neglecting the effect of the circular motion of the row, as well as other sources of drag on the hull like the wave drag. 
{Most importantly, the maximal theoretical boat velocity is reached  for $\alpha = 0$, while real rows have $\alpha \simeq 2.2$.} The stroke duration is also off compared to the theory (Fig.~\ref{opt_reel}(a)). 
As mentioned above, physiology imposes a limit to our model.  Indeed, the mechanical optima identified here are not always attainable by the athletes. In particular, the rowers are not able to hold the pace and row at too high frequencies (or equivalently too small stroke durations).
A given rower should thus choose the smallest possible rows corresponding to the minimal stroke duration he is able to achieve while deploying a maximal force. \\

In addition, there is a physiological relationship between the force exerted by a muscle and its characteristic speed, as shown by Hill \cite{wilkie1949relation,cohen2015weightlifting}. When the speed increases the force decreases, implying that there exists an optimum power developed by the muscle. In the specific case of rowing, the movement is quite complex and involves a lot of different muscle groups. A physiological study would allow to find the optimum for a given athlete and to confront it with the mechanical optimum to choose the oars.

Other physiological and practical aspects can be  important when it comes to the choice of the row length. With smaller rows, the rower would have to raise much more the hands, which is not optimal to pull the row. The techniques for the catch (or blade entry in water) and the release (blade going out of water) should also be changed to adapt with the new oars. 
Furthermore, for a good synchronisation between rowers, it is necessary that all the rowers deploy the same force and have the same row characteristics \cite{boucher2017row}. \\
  
{\red }

\section{Concluding remarks}

The present study deals with the question of optimal oar characteristics with a constant imposed force to model the rower, in contrast with most previous works which considered imposed kinematics. 
This assumption closes the mechanical problem, setting the movement of the rower and the stroke duration. Our theoretical model was validated experimentally in static and dynamic using a robot rowing boat at constant force. We distinguish two regimes depending on whether the force on the blade is dominated by added mass or by pressure drag.
We found that real rowing lies at the cross-over between these two regimes.

The optimal row length and blade size depend on the adopted strategy. If one wants to go as fast as possible without paying attention to the energy consumed (sprint strategy), it is better to use short oars and large blades.  If one however aims at minimising the injected mean power (endurance strategy), long oars and small blade are optimal. Note that olympic rowing races correspond rather to the sprint regime (race duration is around 6 min) and thus the oars should be small while ensuring a reasonable stroke frequency. This is actually the tendency observed historically on rowing (Fig.~\ref{evol}).

To conclude, let us underline that our study aims at providing the key ingredients to perform the optimisation of oar length and blade size for a given rower depending on the rowing category. 
To be more quantitative, the effect of the row angle with respect to the direction of motion, the lift on the blade and the wave drag on the hull could be taken into account.
A more realistic force profile could also be injected in the dynamical equations, which would then have to be solved numerically.
Finally, note that our work can easily be extended to other sports or propulsive mechanisms, such as kayaking, canoeing \footnote{In kayaking, as there is no rowlock, $\alpha = 1$, and  the stroke frequency observed in competitions is much higher than that of rowing (near $100$ strokes per minute), which follows the tendency that the stroke frequency decreases with $\alpha$ (Fig.~\ref{figtorture}(b3)).
Kayak blades have another specific feature: they are very hollow to increase added mass.
However, in comparison with rowing, sprint kayaks tend to go slower as only the rower's arms work (as opposed to legs, back and arms in rowing).} and swimming \cite{pendergast2003energy}. 

\ack{We thank Thomas Baroukh, Renan Cuzon, Fran\c cois Gallaire, Emmanuel Hoang, Kevin Lippera, Hugo Maciejewski  and  Marc Rabaud for fruitful discussions. We thank Augustin Mouterde, and Edouard Jonville for providing the handle force data.}



%
%
%

\section*{References}

\bibliographystyle{unsrt}
\bibliography{article_Optimal_oar}

\end{document}